\documentclass[11pt,a4paper,fleqn]{article}
\begin{document}
\pagestyle{plain}
\newcommand{\be}{\begin{equation}}
\newcommand{\ee}{\end{equation}}
\newcommand{\bea}{\begin{eqnarray}}
\newcommand{\eea}{\end{eqnarray}}
\newcommand{\vp}{\varphi}
\newcommand{\pr}{\prime}
\newcommand{\sech} {{\rm sech}}
\newcommand{\cosech} {{\rm cosech}}
\newcommand{\psib} {\bar{\psi}}
\newcommand{\cosec} {{\rm cosec}}
\def\vs {\vskip .3 true cm}
\centerline{\huge Off-diagonal long-range order in}
\centerline{\huge a one-dimensional many-body problem}
\vskip .5 true cm
\centerline{Guy Auberson \footnote{auberson@lpm.univ-montp2.fr}}
\centerline{Physique Math\'{e}matique et Th\'{e}orique, UMR 5825-CNRS}
\centerline{Universit\'{e} de Montpellier, Montpellier, France}
\vskip .5 true cm
\centerline{Sudhir R. Jain \footnote{srjain@apsara.barc.ernet.in}}
\centerline{Theoretical Physics Division}
\centerline{Bhabha Atomic Research Centre, Trombay, Mumbai 400085, India}
\vskip .5 true cm
\centerline{Avinash Khare \footnote{khare@iopb.res.in}}
\centerline{Institute of Physics, Sachivalaya Marg}
\centerline{Bhubaneswar 751005, Orissa, India}

\date{}

\vskip 0.5 truecm
\noindent
{\bf Abstract}

We prove the existence of
an off-diagonal long-range order in a one-dimensional many-body
problem.

\newpage
It is well known that in a many body system of bosons or fermions it is
possible to have
an off-diagonal long-range order (ODLRO) of the reduced
density matrices in coordinate space representation \cite{yang}.
The onset of ODLRO points at quantum phases and quantum phase
transitions in a many-body system.
For a many-body bosonic system, the
well-known example is the Bose-Einstein condensation (BEC).
It is reasonable to assume that
superfluid He II and superconductors are quantum phases
characterized by the existence of such an order.
A generalized
criterion developed by Penrose and Onsager states that a system exhibits
ODLRO if the largest eigenvalue of the one-particle reduced density matrix
is extensive \cite{penrose-onsager}. Besides these well-known examples of
three-dimensional systems, it was relatively recently shown that there
exists a novel type of ODLRO in the fractional quantum Hall effect ground
state \cite{girvin-macdonald}. Related to this, presence of ODLRO in the
ground state of a Calogero-type model in two dimensions has also been
demonstrated \cite{khare-ray}.

Recently, two of us (JK) \cite{jain-khare} have presented a many-body
system in one dimension with upto next-to-nearest neighbour interaction
and obtained its exact ground state.
For the version of the model where $N$ particles reside
on a circle, they calculated the one and two point correlation functions by
using this ground state and showed that there is no long range order.
Further, by applying the Penrose-Onsager criterion
it was claimed that there is no ODLRO as well.
Unfortunately, the Hamiltonian employed there to describe the system is
not a symmetrised one and since the Penrose-Onsager criterion
does not apply for the case of distinguishable particles, the claim regarding
the absence of ODLRO is incorrect. In this Letter,
we present the symmetrised version
of the many-body system and prove that indeed there exists ODLRO. To
the best of our
knowledge, this is the first example of a one-dimensional system to
possess ODLRO and hence,  quantum phases.

Consider $N$ particles on a circle, described by the symmetrised Hamiltonian,
\bea\label{1}
H &=& - \frac{1}{2} \sum^N_{i=1}\frac{\partial^2}{\partial
  x_i^2} + \sum_{P \in S_N} \nonumber \\
&& \Theta (x_{P(1)}-x_{P(2)})... \Theta
  (x_{P(N-1)}-x_{P(N)}) W (x_{P(1)},...,x_{P(N)}) \, ,
\eea
where $\Theta$ is the step function and $W(x_1,...,x_N)$ is the
$N$-body potential
\bea\label{2}
W(x_1, \cdots ,x_N) &=& g\frac{\pi ^2}{L^2}\sum_{i=1}^{N} \frac{1}{
\sin ^2[{\pi \over L}(x_i-x_{i+1})]} \nonumber \\
&-& G\frac{\pi ^2}{L^2}\sum_{i=1}^{N}
\cot \left[{\pi \over L}(x_{i-1} - x_i) \right]
\cot \left[{\pi \over L}(x_{i} - x_{i+1}) \right].
\eea
In (\ref{1}), the sum is over all permutations on $N$ symbols in the
symmetric group, $S_N$. The exact ground state of the un-symmetrised version
of (\ref{1}) was obtained in \cite{jain-khare}.
Relying on the ground state found there,
we introduce the symmetrized (un-normalized) wave function :
\be\label{3}
\psi_N(x_1,...,x_N) = \phi_N (x_{P(1)},... , x_{P(N)}) \, ,
\ee
where $P$ is the permutation in $S_N$ such that $1> x_{P(1)} > x_{P(2)} >... >
 x_{P(N)} > 0,~ $ and $\phi_N$ is the (un-symmetrized)
exact ground state wave function of the
un-normalized Hamiltonian which is given by
\be\label{4}
\phi_N (x_1,...,x_N) = \prod^N_{n=1} \mid \sin \pi (x_n-x_{n+1})\mid^{\beta} \, ;
 \ \ (x_{N+1}=x_1) \, ,
\ee
provided $g=\beta (\beta - 1),~G=\beta ^2$
(we have set the scale factor $L$ equal to 1). Primitively, the
function (\ref{3}) is defined on the hypercube $[0,1]^N$.

Now for $\beta \ge 2$ it is easily verified that
$\psi_N$ can be continued to a
 multi-periodic function in the whole space ${\cal R}^{N}$ (or
 equivalently on the torus $T^N$):
\be\label{5}
\psi_N(x_1,..., x_{i}+1,...,x_N) = \psi_N (x_1,...,x_i,...,x_N) \, ; \ \
(i = 1,...,N) \, ,
\ee
which belongs to $C^2$ (i.e. is twice continuously
differentiable). Owing to this property and the results of \cite{jain-khare},
$\psi_N$ then obeys the Schr\"odinger equation (with Hamiltonian (1) and with
the same ground state energy $E_0=N(\beta \pi)^2$)
not only in the sector $x_1 > x_2 > ...
> x_N$ but
everywhere. Thus, $\psi_N$ describes the ground state wave function
of the $N$-boson system. Moreover, it is translation-invariant (on
${\cal R}^{N}$):
\be\label{6}
\psi_N (x_1+a,x_2+a,..., x_N+a) = \psi_N (x_1,x_2,...,x_N) \, ; \ \
\forall \ a \ \in \ \cal{R} \, .
\ee

We are interested in the one-particle reduced density matrix, given by
\bea\label{7}
\rho_{N} (x-x') &=& \frac{N}{C_N}\int^1_0 dx_1...\int^1_0 dx_{N-1}
\nonumber \\
&&\psi_N (x_1,..., x_{N-1}, x) \psi_N (x_1,...,x_{N-1}, x') \, ,
\eea
where $C_N$ stands for the squared norm of the wave function:
\be\label{8}
C_N= \int^1_0 dx_1 ...\int^1_0 dx_N \mid \psi_N (x_1,...,x_N)\mid^2 \, .
\ee
That the right hand side (RHS) of Eq. (\ref{7})
defines a (periodic) function of $(x-x')$
is an easy consequence of Eqs. (\ref{5}) and (\ref{6}). The normalization
of $\rho_N$ is such that $\rho_N (0)=N$, the particle density. Further, the
function $\rho_N (\xi)$ is manifestly of positive type on the $U(1)$ group,
which implies that its Fourier coefficients,
\be\label{9}
\rho^{(n)}_N = \int^1_0 d\xi e^{-2i\pi n\xi} \rho_N(\xi) \, ; \ \
(n=0, \pm 1,\pm 2,...) \, ,
\ee
are non-negative (Bochner's theorem). In fact, this directly appears if one
writes their explicit expression
in the form (obtained by using the periodicity property):
\be\label{10}
\rho^{(n)}_N = \frac{N}{C_N}\int^1_0 dx_1 ... \int^1_0 dx_{N-1}
\; \vline  \int^1_0 dx e^{2 i\pi n x}\psi_N(x_1,...,x_{N-1},x)
\vline ^2 \, .
\ee
Since the function $\rho_N$ is not only of
positive type but also {\bf positive} (like $\psi_N$), Eq. (\ref{9})
shows us that
\be\label{11}
\rho^{(0)}_N \geq \rho^{(n)}_N \, ; \ \ (n = \pm 1,\pm 2,...) \, .
\ee

Notice that the coefficients
$\rho_N^{(n)}$, which physically represent the expectation values of the
number of particles having momentum $k_n = 2\pi n$ in the ground state, are
nothing but the eigenvalues of the one-particle reduced density matrix
(diagonal in the $k_n$ representation). According to the Penrose-Onsager
criterion \cite{penrose-onsager}, no condensation can occur in the system
(at least for Bose
particles) if the largest of these eigenvalues is not an extensive quantity
in the thermodynamic limit, that is, if
\be\label{12}
\lim_{N\rightarrow\infty} \frac{\rho^{(0)}_N}{N} = 0 \, .
\ee

Accordingly, we have to evaluate
\be\label{13}
\frac{\rho^{(0)}_N}{N} = \frac{A_N}{C_N} \, ,
\ee
where $C_N$ is given by Eq. (\ref{8}) and
\bea\label{14}
A_N &=& \int^1_0 dx_1...\int^1_0 dx_{N-1} \psi_N (x_1,...,x_{N-1},
0) \nonumber \\
&&\int^1_0 dx \psi_N (x_1,...,x_{N-1}, x).
\eea
Because of the special form, (\ref{3})-(\ref{4}), of the wave
function, the computation of the squared norm $C_N$ is more difficult
than in the  case of $N$ free, impenetrable
particles. As a consequence, the (mainly algebraic) method introduced
long ago by Lenard [26] to deal with the latter case does not apply
here, and we have to resort to another device. For conciseness, we
introduce the notation:
\be\label{15}
S (x_n - x_{n+1}) _ = \mid \sin \pi (x_n- x_{n+1})\mid^{\beta} \, ,
\ee
and define:
\be\label{16}
 S_2 (\triangle) = \int^{\triangle}_0 dx S(x) S(\triangle-x) \, ; \ \ (0\leq
\triangle\leq 1) \, .
\ee
Our starting point will be the following representations of $C_N$ and
$A_N$:
\be\label{17}
C_N = (N-1)! \frac{1}{2\pi} \int^{\infty}_{-\infty} dx e^{-ix}\tilde
{F} (x)^N \, ,
\ee
\be\label{18}
 A_N = (N-1)! \frac{1}{2\pi} \int^{\infty}_{-\infty} dx e^{-ix}\tilde
{F}(x)^{N-3} \bigg [\tilde {F}(x)\tilde {G}(x)
+(N-2)\tilde {H}(x)^2 \bigg ] \, ,
\ee
where
\bea\label{19}
\tilde{F} (x) & = & \int^1_0 d\triangle e^{i\triangle x} S(\triangle)^2 \, ,
\nonumber \\
\tilde{G} (x) & = & \int^1_0 d\triangle e^{i\triangle x} S_2(\triangle)^2 \, ,
\nonumber \\
\tilde{H} (x) & = & \int^1_0 d\triangle e^{i\triangle x} S(\triangle)
S_2 (\triangle) \, .
\eea
The representations (\ref{17})-(\ref{19}) follow from the
convolution structure of
the expressions (\ref{8}) and (\ref{14}) of $C_N$ and $A_N$, when
written in terms
of appropriate variables. Their proof will be given elsewhere \cite{ajk}. Our
aim is to extract from them the large $N$ behaviour of $C_N$ and
$A_N$. Their form is especially suited for this purpose, because  the
integrands in Eqs. (\ref{17}) and (\ref{18}) are entire functions,
as polynomial
combinations of Fourier transforms of functions with compact support
(Eq. (\ref{19})). Indeed, we are then allowed to, first, shift the integrand
and then apply the residue theorem to meromorphic pieces of the
integrands. However, it turns out that the calculations needed for
arbitrary (integer) values of $\beta$ are quite cumbersome. So, in
order to keep the argument clear enough, we shall content ourselves to
present below these calculations in the simplest case, namely $\beta$
= 1 (recall that, strictly speaking, this value is not allowed), being
understood that similar results are obtained for all integers $\beta
\ge 2$. To be sure, after the illustration of the calculation for the
case of $\beta = 1$, we give the final results for arbitrary integer value
of $\beta $.

For $\beta = 1$, $S(\triangle) = \sin \pi\triangle $, and Eq. (\ref{19})
gives, after reductions:
\bea\label{21}
\tilde {F} (x) & = & \frac{2\pi^2}{i}\frac{1-e^{ix}}{x(x^2-4\pi^2)} \, ,
\nonumber \\
\tilde {G} (x) & = & \frac {4\pi^4}{i} \frac{5x^2-4\pi^2}{x^3(x^2-4\pi^2)^3}
+ e^{ix} R^{(-1)}(x) \, , \nonumber \\
\tilde {H} (x) & = & -\frac {4\pi^3}{i}\frac {1}{x(x^2-4\pi^2)^2}
+ e^{ix} R^{(-2)}(x) \, ,
\eea
where $R^{(n)}(x)$ is a generic notation for rational functions
behaving like $x^n$ when $x\rightarrow \infty$, and the precise form
of which will be eventually of no importance. This produces, for the
functions to be integrated in Eqs. (\ref{17}) and (\ref{18}):
\bea\label{22}
&&\tilde {F} (x)^N = \left(\frac{2\pi^2}{i}\right)^{N} \bigg [\frac {1}{[x(x^2-4\pi^2)]^N} +
\sum^N_{n=1} e^{inx} R_n^{(-3N)} (x) \bigg ],\\
&& \tilde {F} (x)^{N-3} [ \tilde {F} (x)\tilde {G} (x)
+(N-2) \tilde {H} (x)^2 ]\nonumber \\
&& =i
\left(\frac {2\pi^2}{i}\right)^N \bigg \{\frac {(2N+1)x^2-4\pi^2}
{[x(x^2-4\pi^2)]^{N+1}}
+ \sum^{N+1}_{n=1} e^{inx} R_n^{(-3N-1)} (x) \bigg \}
\, .
\eea
Let us stress again that these functions, when analytically continued,
are holomorphic in the whole complex plane (the poles appearing in the
first term are exactly cancelled by the remaining ones).

We consider first $C_N$, now given by
\bea\label{24}
C_N &=& (N-1)! \left(\frac{2\pi^2}{i}\right)^N
\nonumber \\
&&\frac {1}{2\pi}
\int^{\infty}_{-\infty} dx e^{-ix} \bigg \{ \frac{ 1}{[x(x^2-4\pi^2)]^N} +
\sum^N_{n=1} e^{inx} R_n^{(-3N)} (x) \bigg \} \, .
\eea
Since the function within the curly bracket is an entire one, we
can shift the integration path to $I \equiv \{ z = x+ia \ \
\mid x \ \varepsilon \ \cal{R} \}$.
Let us choose $a > 0$. Then, by Cauchy theorem
\be
\int_I dz e^{-iz}\sum^N_{n=1} e^{inz} R_n^{(-3N)} (z) = 0 \, .
\ee
Indeed, the integrand is holomorphic above $I$ and is bounded there by
const. $\mid z \mid^{-3N}$, which allows us to close the integration
path at infinity in the {\bf upper} complex plane. We end up with
\be\label{25}
C_N= (N-1)! \left(\frac{2\pi^2}{i}\right)^N
\frac {1}{2\pi}\int_I dz \frac
{e^{-iz}} {z^N(z^2-4\pi^2)^N} \, .
\ee
Similarly, we are allowed to close the integration path at infinity in
Eq. (\ref{25}), but this time in the {\bf lower} complex plane.
For obtaining the large-$N$ asymptotics, we write
\bea\label{26}
& \int_I dz \frac{e^{-iz}}{z^N(z^2-4\pi^2)^N} = \frac {1} {(N-1)!}
\frac {d^{N-1}}{d\alpha^{N-1}}\mid_{\alpha=4\pi^2} \int_I dz
\frac {e^{-iz}}{z^N(z^2-\alpha)} \nonumber \\
& = \frac {-2i\pi} {(N-1)!}
\frac {d^{N-1}}{d\alpha^{N-1}}\mid_{\alpha=4\pi^2} [
R_{+} (\alpha) +R_{-} (\alpha) +R_{0} (\alpha)] \, ,
\eea
where $R_{\pm} (\alpha)$  and $R_0 (\alpha)$ are the residues of the last
integrand at $z = \pm \sqrt{\alpha} $ and $ z = 0$ respectively. They
are readily computed and sum up to
\be\label{28}
R_{+}(\alpha)+R_{-}(\alpha)+R_{0}(\alpha) = (-1)^{M+1}\sum^{\infty}_{s=0}
\frac {(-1)^s} {(2M+2s+2)!} \alpha^s
\ee
for $N=2M+1$.

Using  Eqs. (\ref{25}), (\ref{26}) and (\ref{28}) we then obtain
\bea\label{29}
C_N & = & \left(\frac {2\pi^2}{i}\right)^N (-1)^{M+1}(-i)
\frac{d^{N-1}}{d\alpha^{N-1}}\mid_{\alpha=4\pi^2} \sum^{\infty}_{s=0}\frac
{(-1)^s}{(2M+2s+2)!}\alpha^s \nonumber \\
    & = & (2\pi^2)^N\sum^{\infty}_{n=0} \frac{(N+n-1)!}{n!(3N+2n-1)!}
(-4\pi^2)^n \, .
\eea
The result is exactly the same for even $N$. It suffices now to observe
that the last series alternates in sign and is decreasing to
deduce
\be\label{30}
C_N = (2\pi^2)^N \frac{(N-1)!}{(3N-1)!} \left[ 1+O(\frac{1}{N})\right] \, .
\ee

Our procedure for evaluating $A_N$ is quite similar, and we give below
the final result, viz.,
\be\label{31}
A_N = (2N+1)(2\pi^2)^N \frac {(N-1)!}{(3N)!} \left[1+O(\frac {1}{N} \right] \, .
\ee
Finally, using  Eqs. (\ref{13}), (\ref{30}) and  (\ref{31}),
we obtain
\be\label{32}
\frac{\rho^{(0)}_N}{N} = \frac{2}{3} \left[1+O(\frac{1}{N})\right] \, .
\ee
The same procedure applies for all integer values of $\beta$,
although the algebra
becomes quite involved. The general result (for any
integer $\beta$) is:
\be\label{33}
\lim_{N \rightarrow \infty} \frac{\rho^{(0)}_N}{N}
= \frac{(\beta!)^4 [(3\beta+1)!]^2}{[(2\beta)!]^2[(2\beta+1)!]^3} \, .
\ee
Our method does not generalize in a straightforward manner
to the case of non-integer values of $\beta$, but there is clearly
no reason to expect a different outcome for such
intermediate values.

To summarize, the Penrose-Onsager criterion (\ref{12}) is {\bf not} met
for bosons in this model as the limit of the ratio in (\ref{33}) is
non-zero. The onset of such an order leads to a new thermodynamic phase
of the system.
Thus, we reach the remarkable conclusion that Bose-Einstein condensation is
indeed {\bf possible} in the bosonic version of the $N$-body model discussed
in \cite{jain-khare}.
We recall once again that, to the best of
our knowledge, this is the only example
in one-dimensional statistical mechanics where there exists ODLRO
and hence quantum phases.

{\bf Acknowledgements}

SRJ acknowledges the warm hospitality of the
Institute of Physics, Bhubaneswar
where this work was initiated while AK would like to thank members of the
Laboratoire de Physique Math\'ematique of Montpellier University for warm
hospitality during his trip there as a part of Indo-French collaboration
Project 1501-1502.

\end{document}